\documentclass[usenatbib]{mn2e}
\usepackage{amsmath,amssymb}
\usepackage{graphicx}
\usepackage{subfigure}
\usepackage{times}

\newcommand{\Om}{$\Omega_\textrm m$}
\newcommand{\sig}{$\sigma_8$}
\newcommand{\sigOm}{$\sigma_8(\Omega_\textrm m/0.27)^{0.3}$}
\newcommand{\magneticum}{\textit{Magneticum}}
\newcommand{\planck}{\textit{Planck}}
\newcommand{\Msun}{M_\odot}

\newcommand{\Munich}{$^1$}
\newcommand{\ExcellenceCluster}{$^2$}
\newcommand{\MPA}{$^3$}
\newcommand{\MPE}{$^4$}

\author[S. Bocquet, A. Saro, K. Dolag and J. J. Mohr]
{Sebastian Bocquet\Munich$^,$\ExcellenceCluster\thanks{E-mail: bocquet@usm.lmu.de},
Alex Saro\Munich$^,$\ExcellenceCluster,
Klaus Dolag\Munich$^,$\ExcellenceCluster$^,$\MPA
~and Joseph J. Mohr\Munich$^,$\ExcellenceCluster$^,$\MPE
\\
\Munich Faculty of Physics, Ludwig-Maximilians University, Scheinerstr.\ 1, 81679 Munich, Germany \\
\ExcellenceCluster Excellence Cluster Universe, Boltzmannstr.\ 2, 85748 Garching, Germany \\
\MPA Max Planck Institute for Astrophysics, Karl-Schwarzschild-Str. 1,\ 85748 Garching, Germany \\
\MPE Max Planck Institute for Extraterrestrial Physics, Giessenbachstr. 1,\ 85748 Garching, Germany
}

\title[The halo mass function]{Halo mass function: Baryon impact, fitting formulae\\ and implications for cluster cosmology}

\begin{document}

\maketitle

\begin{abstract}

We use a set of hydrodynamical (Hydro) and dark matter only (DMonly) simulations to calibrate the halo mass function (HMF). We explore the impact of baryons, propose an improved parametrization for spherical overdensity masses and identify differences between our DMonly HMF and previously published HMFs. We use the \textit{Magneticum} simulations, which are well suited because of their accurate treatment of baryons, high resolution, and large cosmological volumes of up to $(3818~\textrm{Mpc})^3$.  Baryonic effects globally decrease the masses of galaxy clusters, which, at a given mass, results in a decrease of their number density. This effect vanishes at high redshift $z\sim2$ and for high masses $M_{200\textrm m}\gtrsim10^{14}M\odot$. We perform cosmological analyses of three idealized approximations to the cluster surveys by the South Pole Telescope (SPT), \textit{Planck}, and eROSITA. We pursue two main questions: (1) What is the impact of baryons? -- For the SPT-like and the \textit{Planck}-like samples, the impact of baryons on cosmological results is negligible. In the eROSITA-like case, however, neglecting the baryonic impact leads to an underestimate of $\Omega_\textrm m$ by about $0.01$, which is comparable to the expected uncertainty from eROSITA. (2) How does our DMonly HMF compare with previous work? -- For the \textit{Planck}-like sample, results obtained using our DMonly HMF are shifted by $\Delta(\sigma_8)\simeq\Delta(\sigma_8(\Omega_\textrm m/0.27)^{0.3})\simeq0.02$ with respect to results obtained using the Tinker et al. (2008) fit. This suggests that using our HMF would shift results from \textit{Planck} clusters toward better agreement with CMB anisotropy measurements. Finally, we discuss biases that can be introduced through inadequate HMF parametrizations that introduce false cosmological sensitivity.

\end{abstract}


%%%%%%%%%%%%%%%%%%%%%%%%%%%%%%%%%%%%%%%%%%%%%%%%

\section{Introduction}

Galaxy clusters are the largest collapsed objects in the Universe. Their distribution in mass and redshift is highly sensitive to key cosmological parameters such as the matter density \Om, or the amount of matter fluctuations in the Universe \sig\ \citep[e.g.][]{henry91,white93b}. Furthermore, they can be used to constrain models of dark energy, the cosmic growth rate, and the neutrino sector \citep{wang98,haiman01}. Catalogues from different cluster surveys have proven to be useful cosmological probes \citep[e.g.][]{vikhlinin09b, mantz10b, rozo10, benson13,hasselfield13, planck15-24, bocquet15,wtg14d}.

The predicted abundance of galaxy clusters is linked to the linear matter power spectrum through the halo mass function (HMF), which was first estimated analytically \citep{press74}. Since then, numerical $N$-body simulations have been used to calibrate fitting functions \citep[e.g.][]{jenkins01, sheth01, white02, reed03, warren06, lukic07, reed07, tinker08, crocce10, bhattacharya11, courtin11, angulo12, watson13}. Most of the above studies focus on the friends-of-friends (FoF) halo definition \citep{davis85}. However, real cluster samples are typically defined in terms of spherical overdensity (SO) masses. Only very few HMFs exist for different overdensity definitions \citep[e.g.][hereafter Tinker08, Watson13]{tinker08,watson13}, and the former has developed into the standard reference used in most cluster cosmology analyses.

For a particular HMF parametrization to be useful in cosmological studies, it is crucially important that it correctly captures both the variation in the HMF with redshift, and the sensitivity to cosmological parameters of interest such as the matter density \Om, the dark energy density $\Omega_\Lambda$, the dark energy equation of state parameter $w$, and \sig. An ideal situation would be for the HMF shape parametrization to be universal, where the variation with cosmology would be entirely captured by the cosmological sensitivity of the linear power spectrum of density fluctuations.  For a FoF halo definition with linking length $b=0.2$, or SO $\Delta_{180\textrm{mean}}$, the HMF has been found to be approximately universal over a wide range of redshifts and cosmologies \citep{jenkins01}. More recently, \cite{bhattacharya11} have used a set of $\Lambda$CDM and $w$CDM simulations to investigate the dependence of the FoF HMF on cosmological parameters. Their fit is accurate to 2\% for $\Lambda$CDM, and it describes the $w$CDM HMF to within 10\%. Similar results have also been reported in \cite{courtin11}, although with slightly larger uncertainties.

Any HMF obtained from $N$-body dark matter only simulations potentially suffers from some bias because the baryonic component of the clusters is neglected. Recently, various authors have investigated the baryonic impact on the halo HMF using hydrodynamic simulations \citep[e.g.][]{cui12, cui14, cusworth14, martizzi14, velliscig14, vogelsberger14, schaller15}. Their conclusions are highly sensitive to the details of the treatment of the baryonic component. For example, models without feedback from active galactic nuclei (AGN) lead to higher cluster masses (or higher abundance at fixed mass) than dark matter only simulations. Adding AGN feedback, however, leads to HMFs that are up to 20\% lower than, or about equal to the dark matter only prediction. Also, these baryonic effects are stronger for low cluster masses, and hardly affect the high-mass population. These shifts in the predicted HMFs nearly correspond to the level of uncertainty from current cluster abundance measurements. Therefore, studies of the baryonic impact on the HMF are extremely important for progress in cluster cosmology.

In this work, we use haloes extracted from the hydrodynamical \magneticum\ simulations \citep[Dolag et al., in prep.; see also][]{hirschmann14,saro14,dolag15,teklu15}. In combination with their dark matter only counterparts, these simulations allow us to investigate several key aspects of the HMF. In particular, we focus on the impact of baryons, universality for various SO definitions, and a comparison of our dark matter only HMF with previously published fits. In Section~\ref{sec:data}, we present the \magneticum\ simulations and describe how the cluster samples are extracted. We discuss theoretical aspects of the HMF in Section~\ref{sec:analysis}, where we also present a novel approach for parametrizing the HMF for SO different from $\Delta_{200\textrm m}$. In Section~\ref{sec:results}, we present our HMFs for different SO definitions, discuss the baryon impact and investigate differences between our dark matter only HMF and previous work. The cosmological impact of all these effects is presented in Section~\ref{sec:cosmoimpact}. We summarize our results in Section~\ref{sec:summary}, where we also present step-by-step instructions on how to use our HMF fitting formulae.

We consider the following SO definitions: (1) \textit{mean overdensity} mass $M_{200\textrm m}$, which is the mass enclosed within a sphere of radius $r_{200\textrm m}$, in which the mean enclosed matter density is equal to 200 times the mean universal matter density $\bar\rho_\textrm m (z)$ at the cluster's redshift, and (2) \textit{critical overdensity} masses $M_{500\textrm c}$ ($M_{200\textrm c}$), which are analogous to (1) but enclosed within $r_{500\textrm c}$ ($r_{200\textrm c}$), and defined with respect to the critical density $\rho_\textrm{crit} (z)$. The corresponding overdensities are $\Delta_{200\textrm m}$, $\Delta_{500\textrm c}$ and $\Delta_{200\textrm c}$. The critical density is $\rho_\textrm{crit}(z) = 3H^2(z)/8\pi G$, where $H(z)$ is the Hubble parameter. The mean matter density is $\bar\rho_\textrm m (z) = \Omega_\textrm m (z)\rho_\textrm{crit}(z)$ with $\Omega_\textrm m (z) = \Omega_\textrm m (1+z)^3/E^2(z)$, and where $E(z)\equiv H(z)/H_0$.

%%%%%%%%%%%%%%%%%%%%%%%%%%%%%%%%%%%%%%%%%%%%%%%%

\begin{table*}
\begin{minipage}{\textwidth}
\centering
\caption{Boxes of the \magneticum\ simulations used in this work. The number of haloes $N_{(z=0)}$ refer to the Hydro runs, and $M_{200\textrm m}$.}
\begin{tabular}{lcccccccccc}
Box & Size $L_\textrm{box}$ & \multicolumn{3}{c}{grav. softening length (kpc)} & $N_\textrm{particles}$ & $m_\textrm{DMparticle}$  & $M_\textrm{halo, min}$ & \multicolumn{2}{c}{Simulation} &$N_{(z=0)}$  \\ 
&  &DM&gas&stars&& ($\Msun$) & ($\Msun$) &Hydro&DMonly&\\ \hline
4/uhr & $68.1~\textrm{Mpc}$ &$1.4$&$1.4$&$0.7$& $2 \times576^3$ & $5.3\times10^7$ &  $6.2\times10^{11}$ &\checkmark&\checkmark& --\footnote{The Hydro simulation of Box4/uhr is only run to $z=0.13$. Its DMonly counterpart contains 835 haloes at redshift $z=0$.}\\
3/hr & $182~\textrm{Mpc}$ &$3.75$&$3.75$&$2$& $2 \times576^3$ & $9.8\times10^8$ & $1.1\times10^{13}$ & \checkmark&\checkmark&$966$\\
2/hr & $500~\textrm{Mpc}$ &$3.75$&$3.75$&$2$& $2 \times1584^3$ & $9.8\times10^8$ & $1.1\times10^{13}$ & \checkmark&&$21\,528$\\
2b/hr & $909~\textrm{Mpc}$ &$3.75$&$3.75$&$2$& $2 \times2880^3$ & $9.8\times10^8$ & $1.1\times10^{13}$ & \checkmark&&--\footnote{Box2b/hr is only run to $z=0.2$. At this redshift, it contains 109\,578 haloes.}\\
1/mr & $1274~\textrm{Mpc}$ &$10$&$10$&$5$& $2 \times1526^3$ & $1.9\times10^{10}$ & $2.2\times10^{14}$ &\checkmark &\checkmark&$8528$\\
0/mr & $3818~\textrm{Mpc}$ &$10$&$10$&$5$& $2 \times4563^3$ & $1.9\times10^{10}$ & $2.2\times10^{14}$ &\checkmark&& $227\,400$
\end{tabular}
\label{tab:sims}
\end{minipage}
\end{table*}

%%%%%%%%%%

\section{Simulations and Cluster selection} \label{sec:data}

We will refer to our hydrodynamical simulations and HMFs as ``Hydro'', and to their dark matter only counterparts as ``DMonly''.

\subsection{The \magneticum\ simulations}

In this work, we use a subset of the cosmological Hydro and DMonly boxes from the \magneticum\ simulation set (Dolag et al., in prep.) as highlighted in Table~\ref{tab:sims}. The simulations are based on the parallel cosmological TreePM-SPH code {\small P-GADGET3} \citep{springel05gadget}. We use an entropy-conserving formulation of SPH \citep{springel02} and a higher order kernel based on the bias-corrected, sixth-order Wendland kernel \citep{dehnen12} with 295 neighbours, which together with a low-viscosity SPH scheme allows us to properly track turbulence within galaxy clusters \citep{dolag05,donnert13,beck16}. 

We also allow for isotropic thermal conduction with $1/20$ of the classical Spitzer value \citep{dolag04thermalconduction}. The simulation code includes a treatment of radiative cooling, heating from a uniform, time-dependent ultraviolet background and star formation with the associated feedback processes. The latter is based on a sub-resolution model for the multiphase structure of the interstellar medium \citep{springel03multiphase}.

We compute radiative cooling rates following the same procedure as presented by \citet{Wiersma09}, and account for the presence of an evolving ultraviolet background \citep{haardt01}. Contributions to cooling from each element  have been pre-computed using the publicly available {\small CLOUDY} photo-ionisation code \citep{Ferland98} for an optically thin gas in (photo-)ionisation equilibrium.

Our simulations also incorporate a detailed treatment of stellar evolution and chemical enrichment following \citet{tornatore07}, a multiphase model for star-formation \citep{springel03multiphase}, and feedback processes associated with supernovae driven galactic winds and AGN \citep{springel03multiphase,dimatteo08,fabian10}. Additional details about the simulation code are available elsewhere \citep[e.g.][]{hirschmann14}.

Initial conditions are created at redshift $z_\textrm{ini}=60$ applying the Zel'dovich approximation\footnote{We discuss the impact of the initial conditions in Appendix~\ref{sec:initialization}.} for a spatially flat $\Lambda$CDM cosmology with parameter values from WMAP7 \citep{komatsu11}: matter density $\Omega_\textrm m=0.272$, baryon density $\Omega_\textrm b=0.0456$, scalar spectral index $n_s=0.963$, variance in the matter field\footnote{See Equation~\ref{eq:sigma} for the definition.} $\sigma_8=0.809$, and Hubble constant $H_0=70.4$~km~s$^{-1}$~Mpc$^{-1}$. The simulation boxes have periodic boundary conditions and are initially occupied by an equal number of gas and dark matter particles. Their relative masses reflect the global baryon fraction $\Omega_\textrm b/\Omega_\textrm m$. To minimize numerical differences between the Hydro and the DMonly simulations, we set up the DMonly simulations with an equal number of two types of collisionless particles, whose masses are equal to the mass of the dark matter and of the gas particles, respectively, of the corresponding Hydro runs.

%%%%%%%%%%

\begin{figure}
\includegraphics[width=\columnwidth]{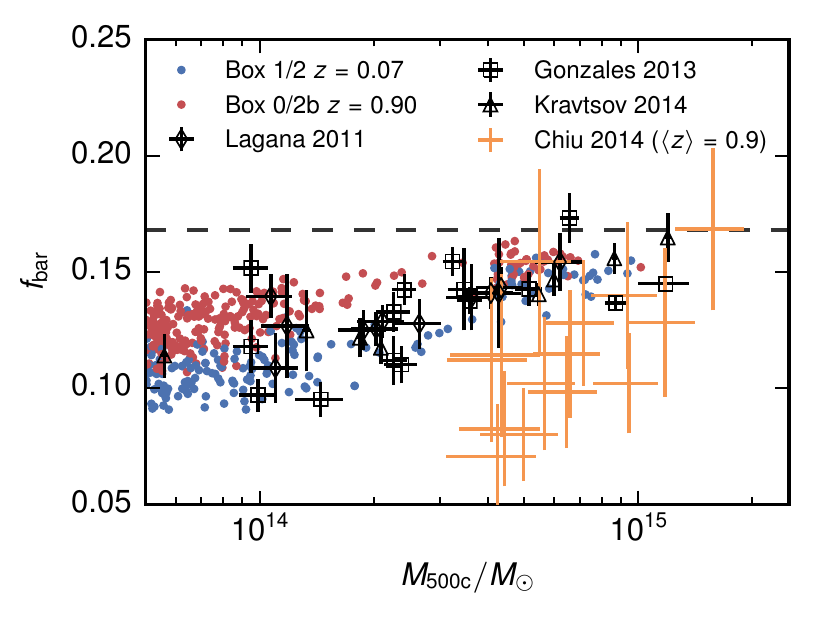}
\caption{Baryon fractions for a subset of haloes extracted from our simulations and for observations at low and high redshift.  As discussed in the text, the offset between the low- and high-redshift observations is largely due to systematic differences in the cluster binding mass measurements.}
\label{fig:f_bar}
\end{figure}

%%%%%%%%%%

The \magneticum\ Hydro simulations have been shown to correctly reproduce AGN luminosity functions \citep{hirschmann14} and cluster pressure profiles \citep{planck13-5,mcdonald14}. In Fig.~\ref{fig:f_bar}, we compare the baryon fraction $f_\textrm{bar}$ of haloes extracted from our simulations with recent observational results on galaxy cluster and group mass scales. Simulation values at $z=0.9$ (red) are higher than the simulation values at $z=0.07$ (blue), indicating that the depletion of the intracluster medium is increasing over time.  Also shown are observational results at low redshift \citep{lagana11,gonzales13}, where we update the total and stellar masses for a subset of clusters as presented in \cite{kravtsov14}, and at high redshift $\left< z \right>=0.9$ \citep{chiu16}. The low redshift observations and simulations are in good agreement, but the simulations exhibit systematically larger baryon fractions than are observed for the sample of clusters at higher redshifts.

The comparison of the observations with the simulations is complicated, given that the observations rely on measured cluster masses.  As noted in \citet{bocquet15}, the X-ray hydrostatic masses tend to be lower in comparison to masses obtained using velocity dispersions or through abundance matching that includes constraints from external cosmological probes;  other studies using weak lensing mass constraints come to similar conclusions \citep[e.g.][]{vonderlinden14,hoekstra15}.  In Fig.~\ref{fig:f_bar} the low redshift baryon fractions rely on X-ray hydrostatic masses, and the high redshift measurements adopt the abundance matching masses referred to above.  In moving from hydrostatic mass to abundance matching mass, there is an increase of $\sim$44\% in the binding mass and a reduction of $\sim$27\% in the baryon fraction \citep{chiu16}.  The observed and simulated baryon fractions are in agreement if we adopt these shifts as the current scale of the systematic uncertainties.

%%%%%%%%%%

\subsection{Halo selection} \label{sec:selection}

Haloes are initially identified through a parallel FoF algorithm with linking length $b=0.16$. The FoF links over dark matter particles only. We then compute SO masses (for $\Delta_{200\textrm m}$, $\Delta_{200\textrm c}$ and $\Delta_{500\textrm c}$) of each halo centered at the deepest potential point with the parallel {\small SUBFIND} algorithm \citep{Springel01,Dolag09sub}. In Appendix~\ref{sec:substructure}, we discuss an alternative approach where subhaloes are identified as individual objects.

To ensure that haloes extracted from the Hydro simulations are not affected by issues related to resolution and numerical artefacts, we apply very conservative convergence criteria. For each box, and for each overdensity $\Delta$, we only consider haloes that contain more than $10^4$ dark matter particles within $r_\Delta$. We then construct catalogues applying the lower mass limits shown in Table \ref{tab:sims}. We further apply an upper mass limit that corresponds to the lower limit of the next larger box, or to $10^{16}\Msun$ for the largest boxes (see also Fig.~\ref{fig:dNdM}). We extract cluster catalogues at seven redshifts\footnote{We use the Hydro run of Box4/uhr for redshifts $z\geq0.13$.} that are roughly equally spaced in cosmic time with $\Delta t\sim1.6$ Gyr. This time step is chosen to be larger than the typical dynamic time of a halo, and we therefore work under the assumption that there is no correlation between the different snapshots.

%%%%%%%%%%%%%%%%%%%%%%%%%%%%%%%%%%%%%%%%%%%%%%%%

\section{Analysis Method} \label{sec:analysis}

We provide the theoretical background on the HMF and introduce the fitting form we adopt. We also present the method used to perform the multi-dimensional fits for the HMF parameters.

%%%%%%%%%%

\subsection{The halo mass function}

The comoving number density of haloes of mass $M$ is
\begin{equation}
\frac{dn}{dM} = f(\sigma) \frac{\bar\rho_\textrm{m}}{M} \frac{d\ln\sigma^{-1}}{dM},
\label{eq:MF}
\end{equation}
with the mean matter density  $\bar\rho_\textrm{m}$ (at redshift $z=0$), and
\begin{equation} \label{eq:sigma}
\sigma^2(M,z) \equiv \frac 1{2\pi^2} \int P(k,z) \hat W^2(kR) k^2 dk,
\end{equation}
which is the variance of the matter density field $P(k,z)$ smoothed with the Fourier transform $\hat W$ of the real-space top-hat window function of radius $R = (3M/4\pi \bar\rho_\textrm{m})^{1/3}$.
The function $f(\sigma)$ is commonly parametrized as
\begin{equation}
f(\sigma) = A\left[ \left(\frac \sigma b \right) ^{-a} +1\right] \exp \left(-\frac c {\sigma^2} \right)
\label{eq:fsigma}
\end{equation}
with four parameters $A,a,b,c$ that need to be calibrated \citep[e.g.][]{jenkins01}. Here, $A$ sets the overall normalization, $a$ and $b$ are the slope and normalization of the low-mass power law, and $c$ sets the scale of a high-mass exponential cutoff.

The fit parameters in $f(\sigma)$ will in general depend on redshift and cosmology, which must be accounted for. In the following we note that this dependence is weak in certain special cases; this is referred to as universality of the HMF.

%%%%%%%%%%

\subsection{Halo mass function for spherical overdensity masses} \label{sec:SO}

Many studies of the HMF are performed using the FoF technique. For a linking length $b\simeq0.2$, the resulting HMF is very close to being universal \citep{jenkins01}. However, for observational reasons, real cluster masses are measured in terms of SO masses.

\subsubsection{The halo mass function for $\Delta_{200\textrm m}$} \label{sec:form200m}

When using a suitable SO $\Delta_\textrm{mean}\sim180$, the above HMF fitting function is still close to being universal \citep{jenkins01}. Similarly, Tinker08 use $\Delta_{200\textrm m}$ as their universal mass definition, and Watson13 argue for $\Delta_{178\textrm m}$. These overdensity definitions are all very similar; we adopt $\Delta_{200\textrm m}$ in this work.

We follow previous work and allow for departures from universality by parametrizing a possible redshift dependence of the HMF in $\Delta_{200\textrm m}$ as a power law in $1+z$:
\begin{align}
A(z) &= A_0 (1+z)^{A_z} \nonumber\\
a(z) &= a_0 (1+z)^{a_z} \nonumber\\
b(z) &= b_0 (1+z)^{b_z} \nonumber\\
c(z) &= c_0 (1+z)^{c_z} \label{eq:MF_z}
\end{align}
where the subscript 0 denotes the values at redshift $z=0$, and where $A_z, a_z, b_z, c_z$ are additional fit parameters. Note that many authors assume the cutoff scale $c$ to be constant ($c_z=0$, e.g. Tinker08, Watson13).

\subsubsection{Toward a universal mass function for $\Delta_{500\textrm c}$ and $\Delta_{200\textrm c}$}

We also wish to calibrate the HMF for $\Delta_{500\textrm c}$, which is a convenient mass definition within X-ray studies of clusters where the emission cannot easily be traced beyond $r_{500\textrm c}$, and for $\Delta_{200\textrm c}$, which is used for measurements of cluster galaxy velocity dispersions and for weak gravitational lensing shear profiles. It is not a priori clear that one can simply use the same form of the fitting function that is valid for $\Delta_{200\textrm m}$, and calibrate it against simulations at other overdensities, as one might miss some redshift and cosmology dependent behaviour. Remember, for example, the very different redshift evolution of $\bar\rho_\textrm m (z)$ and $\rho_\textrm{crit}(z)$. 

Tinker08 provide the HMF for a range of different $\Delta_\textrm{mean}$.  One  then uses these HMFs together with $\Delta_\textrm{mean}(z)=\Delta_\textrm{crit}/\Omega_\textrm m (z)$ to convert from critical to mean density as a function of redshift. Their approach relies on the implicit assumption that the fitting function correctly captures the behaviour for every $\Delta_\textrm{mean}$. Watson13 provide a correction to their $\Delta_{178\textrm m}$ HMF that depends on $\Delta_\textrm{mean}(z)$.

We choose a novel approach, based on the assumption that the HMF is approximately universal in $\Delta_{200\textrm m}$. Using the HMF for  $\Delta_{200\textrm m}$ we then determine the HMF for $\Delta_{500\textrm c}$ as: 
\begin{align} \label{eq:MF500c}
\frac{dn}{dM_{500\textrm c}} &= \frac{dn}{dM_{200\textrm m}} \frac{dM_{200\textrm m}}{dM_{500\textrm c}} \nonumber\\
&= \left(f(\sigma)\frac{\bar\rho_\textrm m}{M_{500\textrm c}}\frac{d\ln\sigma^{-1}} {dM_{500\textrm c}}\right) \frac{M_{500\textrm c}}{M_{200\textrm m}}.
\end{align}
The first factor now has the same functional form as the HMF for $\Delta_{200\textrm m}$ (Equations~\ref{eq:MF}--\ref{eq:MF_z}). The second factor describes the conversion between the two different SO mass definitions. This conversion is discussed below. The crucial point is that the new functional form of the HMF established in Equation~\ref{eq:MF500c} has the same close-to-universal properties as the HMF for $\Delta_{200\textrm m}$. Note that, due to uncertainties and systematic scatter in the $M_{500\textrm c}$-$M_{200\textrm m}$ conversion, $f(\sigma)$ in Equation~\ref{eq:MF500c} still needs to be calibrated against the $\Delta_{500\textrm c}$ simulation data.

SO masses can be converted from one to the other assuming a cluster density profile and a mass-concentration relation. We use the parametrizations by \cite{navarro97} and \cite{duffy08} and establish an analytic fit for $M_{500\textrm c}/M_{200\textrm m}$. This fit for $M_{500\textrm c}/M_{200\textrm m}$ depends on mass, redshift, and \Om\ (which is involved in the overdensity conversion). The following prescription reproduces the mass conversion at the few percent level in the range $0<z<2$, $10^{13}<M_{500\textrm c}/\Msun<10^{16}$, and $0.1<\Omega_\textrm m<0.5$:
\begin{equation} \label{eq:dMvirdM}
\frac{M_{500\textrm c}}{M_{200\textrm m}} \equiv \alpha + \beta \ln \frac{M_{500\textrm c}}{\Msun}.
\end{equation}
The parameters $\alpha$ and $\beta$ are functions of \Om\ and redshift:
\begin{align}
\beta (\Omega_\textrm m) &= -1.70\times10^{-2} + \Omega_\textrm m\; 3.74\times10^{-3} \nonumber\\
\alpha(\Omega_\textrm m,z) &= \alpha_0\frac{\alpha_1\;z+\alpha_2}{z+\alpha_2} \nonumber\\
\alpha_0 (\Omega_\textrm m) &= 0.880 + 0.329\; \Omega_\textrm m \nonumber\\
\alpha_1 (\Omega_\textrm m) &= 1.00 + 4.31\times10^{-2}/\Omega_\textrm m \nonumber\\
\alpha_2 (\Omega_\textrm m) &= -0.365 + 0.254/\Omega_\textrm m \label{eq:lastM500M200}.
\end{align}

In an analogous way, we establish the form of the HMF for $\Delta_{200\textrm c}$ in Appendix~\ref{sec:SO200c}. In Section~\ref{sec:summary} we present step-by-step instructions of how to compute our HMF for either of the three SO definitions. This presentation explicitly shows how the SO mass conversions are used.

%%%%%%%%%%

\begin{figure*}
\includegraphics[width=\hsize]{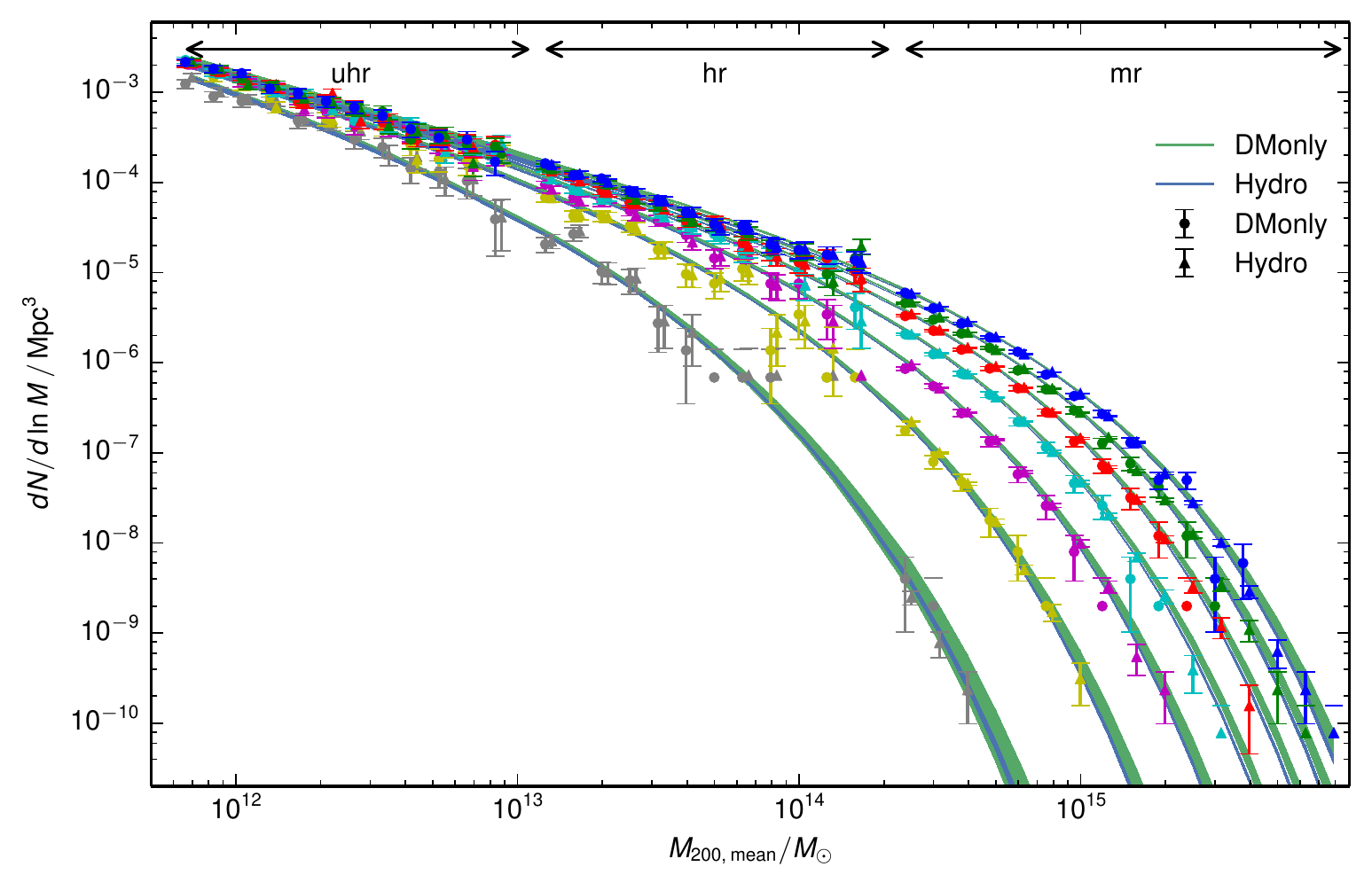}
\caption{Halo mass function $dN/d\ln M$ from our Hydro and DMonly simulations. Redshift is increasing from top to bottom and takes values $z=0,0.13,0.3,0.5,0.8,1.2,2$. The data points are slightly offset in mass for improved readability. We also show the 2$\sigma$ allowed regions of our Hydro (blue region) and DMonly (green region) fits. For Hydro, the error bars and allowed region are narrower because we also use data from the much larger Box0/mr.}
\label{fig:dNdM}
\end{figure*}

%%%%%%%%%%

\subsection{Finite volume correction}

Throughout this work, we use cluster samples produced by simulations to understand the HMF observed in the real Universe. However, there is one subtle difference that needs to be accounted for: in contrast to the Universe, every simulation box is finite in size. Therefore, we can only capture modes in the density field that are smaller than $L_\textrm{box}$. This means that there is an upper mass limit corresponding to the longest modes, beyond which the simulations will systematically underestimate the number of objects.

We correct for this effect following the approach of previous analyses \citep[][Watson13]{lukic07,bhattacharya11}. Briefly, the variance of fluctuations $\sigma(M)$ is corrected by subtracting the variance at scales corresponding to the box size $\sigma(R_\textrm{box})$:
\begin{equation}
\sigma^2_\textrm{corrected}(M) = \sigma^2_\textrm{theory}(M) - \sigma^2(R_\textrm{box}),
\end{equation}
where, for simplicity, we equate the spherical volume $4/3 \pi R_\textrm{box}^3$ to the cubical simulation volume $L_\textrm{box}^3$. However, as we apply upper mass limits to the cluster samples (see Section~\ref{sec:selection}), the correction has negligible impact on our analysis. In fact, for each box size, the correction would become important at masses that are about 2 orders of magnitude larger than the corresponding upper mass limit we apply. We test these finite volume effects by reproducing the results presented in Section~\ref{sec:results} without the correction; in this case, the results do not significantly change. Nevertheless, we apply the correction to each of our boxes.

%%%%%%%%%%

\subsection{Parameter estimation}\label{sec:likelihood}

We use a Bayesian likelihood approach, which allows us to correctly capture the Poisson errors on the measured number of clusters as a function of their mass and redshift. This choice differs from using (Gaussian) $\chi^2$ statistics and jackknife errors (e.g. Tinker08), or corrections to $\chi^2$ statistics to account for the Poisson errors (e.g. Watson13).

The likelihood at each point $\boldsymbol p$ in parameter space is calculated in the following way:  We calculate the matter power spectrum using the transfer function of \cite{eisenstein98,eisenstein99a}, taking baryonic effects into account. This is the same prescription used to set up the initial conditions of the \magneticum\ simulations. We evaluate the likelihood $\mathcal L$ by applying Poisson statistics in log-spaced mass bins of size $\Delta \log_{10}M = 0.1$ \citep{cash79}:
\begin{equation}
\ln \mathcal L = \sum_i \ln\frac{dn(M_i|\boldsymbol p,z_i)}{d M} - \int \frac{dn(M|\boldsymbol p,z)}{dM}dM,
\end{equation}
up to an arbitrary constant, and where $i$ runs over all clusters in the sample. The second term equals the total number of expected clusters. We have checked that decreasing the bin size does not change our results.

In practice, given a set of parameters $\boldsymbol p$, we perform the above calculation for each redshift and for each of the simulation boxes, and sum the log-likelihoods. When fitting for the HMF in this way, we are facing a problem with moderately large dimensionality (8 free parameters); we utilize the \texttt{emcee}\footnote{http://dan.iel.fm/emcee/current} code for efficient exploration of parameter space \citep{foreman-mackey12}. We test our fitting procedure against several mock catalogues that contain a factor 100 times more clusters than our simulation data. In these tests we recover the input values within the statistical uncertainties and conclude that our fitting method is unbiased to a level that is much smaller than the uncertainties we report.

%%%%%%%%%%

\begin{figure*}
\includegraphics[width=\textwidth]{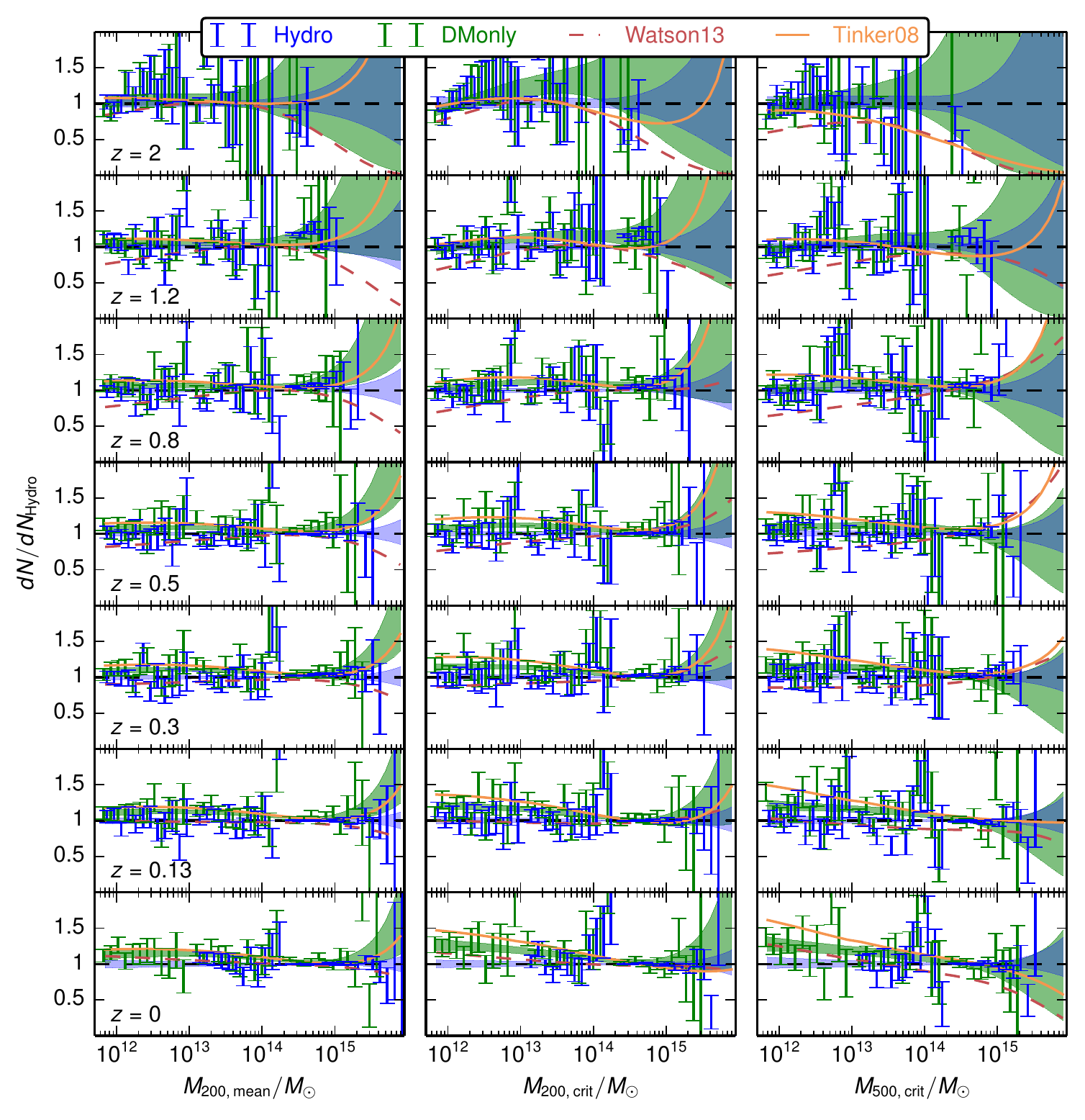}
\caption{Number density of haloes relative to our Hydro fit for $M_{200\textrm m}$ (left panels), $M_{200\textrm c}$ (center panels), and  $M_{500\textrm c}$ (right panels). The coloured bands correspond to the $2\sigma$ allowed regions of our fits; the data points are slightly offset in mass for better readability. For masses above $M>2.2\times 10^{14}\Msun$, the Hydro (DMonly) data points shown here are extracted from Box0 (Box1). The red dashed line shows the fit by \citet{watson13}, and the orange line shows the fit by \citet{tinker08}.}
\label{fig:dNdM_relative}
\end{figure*}

\begin{table*}
\caption{Best-fitting HMF parameters for DMonly (Box1+3+4) and Hydro (Box0+1+3+4) simulations. The covariance matrix for Hydro $M_{500\textrm c}$ is shown in Table~\ref{tab:covmatM500c}. The HMFs for $M_{500\textrm c}$ and $M_{200\textrm c}$ are calculated from Equations~\ref{eq:MF500c} and \ref{eq:MF200c}, respectively.}
\begin{tabular}{lcrrrrrrr}
Parameter & \multicolumn{1}{c}{$A$} & \multicolumn{1}{c}{$a$} & \multicolumn{1}{c}{$b$} & \multicolumn{1}{c}{$c$} & \multicolumn{1}{c}{$A_z$} & \multicolumn{1}{c}{$a_z$} & \multicolumn{1}{c}{$b_z$} & \multicolumn{1}{c}{$c_z$} \\ \hline
\multicolumn{2}{l}{$M_{200\textrm m}$}\\
DMonly & $0.175$ & $1.53$ & $2.55$ & $1.19$ & $-0.012$ & $-0.040$ & $-0.194$ & $-0.021$ \\
Hydro & $0.228$ & $2.15$ & $1.69$ & $1.30$ & $0.285$ & $-0.058$ & $-0.366$ & $-0.045$ \\ \hline
\multicolumn{2}{l}{$M_{200\textrm c}$}\\
DMonly & $0.222$ & $1.71$ & $2.24$ & $1.46$ & $0.269$ & $0.321$ & $-0.621$ & $-0.153$\\
Hydro & $0.202$ & $2.21$ & $2.00$ & $1.57$ & $1.147$ & $0.375$ & $-1.074$ & $-0.196$\\\hline
\multicolumn{2}{l}{$M_{500\textrm c}$}\\
DMonly & $0.241$ & $2.18$ & $2.35$ & $2.02$ & $0.370$ & $0.251$ & $-0.698$ & $-0.310$ \\
Hydro & $0.180$ & $2.29$ & $2.44$ & $1.97$ & $1.088$ & $0.150$ & $-1.008$ & $-0.322$
\end{tabular}
\label{tab:MFparams}
\end{table*}

\begin{table*}
\caption{Covariance matrix for our Hydro HMF in $M_{500\textrm c}$. The corresponding best-fitting values are shown in Table~\ref{tab:MFparams}.}
\begin{tabular}{lrrrrrrrr}
& \multicolumn{1}{c}{$A$} & \multicolumn{1}{c}{$a$} & \multicolumn{1}{c}{$b$} & \multicolumn{1}{c}{$c$} & \multicolumn{1}{c}{$A_z$} & \multicolumn{1}{c}{$a_z$} & \multicolumn{1}{c}{$b_z$} & \multicolumn{1}{c}{$c_z$}\\ \hline
$A$ & $3.33\times10^{-4}$ & $4.58\times10^{-4}$ & $-2.61\times10^{-3}$ & $1.89\times10^{-5}$ & $-3.56\times10^{-3}$ & $-2.25\times10^{-4}$ & $2.22\times10^{-3}$ & $7.38\times10^{-5}$\\
$a$ & & $6.44\times10^{-3}$ & $-7.72\times10^{-3}$ & $1.60\times10^{-3}$ & $3.31\times10^{-3}$ & $2.67\times10^{-5}$ & $-3.55\times10^{-4}$ & $-4.13\times10^{-4}$\\
$b$ & & & $2.39\times10^{-2}$ & $-1.10\times10^{-3}$ & $2.34\times10^{-2}$ & $1.88\times10^{-3}$ & $-1.63\times10^{-2}$ & $-3.66\times10^{-4}$\\
$c$ & & & & $5.12\times10^{-4}$ & $2.71\times10^{-3}$ & $3.01\times10^{-4}$ & $-1.46\times10^{-3}$ & $-2.05\times10^{-4}$\\
$A_z$ & & & & & $1.16\times10^{-1}$ & $1.65\times10^{-2}$ & $-7.71\times10^{-2}$ & $-3.08\times10^{-3}$\\
$a_z$ & & & & & & $4.18\times10^{-3}$ & $-1.24\times10^{-2}$ & $-2.62\times10^{-4}$\\
$b_z$ & & & & & & & $5.33\times10^{-2}$ & $2.10\times10^{-3}$\\
$c_z$ & & & & & & & & $2.40\times10^{-4}$
\end{tabular}
\label{tab:covmatM500c}
\end{table*}

%%%%%%%%%%%%%%%%%%%%%%%%%%%%%%%%%%%%%%%%%%%%%%%%

\section{\textit{magneticum} mass functions} \label{sec:results}

In Fig.~\ref{fig:dNdM}, we show the number density of haloes from our Hydro and DMonly simulations at seven different redshifts. The error bars show the Poisson uncertainty on the measured numbers. We also show the $2\sigma$ allowed regions of the fitting functions. The characteristic low-mass power law and the steepening at high mass are clearly visible.

The impact of baryons can be better distinguished in Fig.~\ref{fig:dNdM_relative}, which shows the number density of clusters from both our simulations relative to the best-fitting function for Hydro. We also show the allowed $2\sigma$ regions of our fits and two external dark matter only fitting functions (Tinker08, Watson13).

\subsection{Impact of baryons} \label{sec:baryonimpact}

Throughout the different SO definitions, the comparison of our Hydro and DMonly simulations tells an interesting story: At $z=2$, the highest redshift we consider, there is essentially no difference between our Hydro and DMonly HMFs. As structure formation continues, baryonic effects become important, and at redshift $z\sim1$, we observe that baryon depletion is important for low-mass clusters up to $\sim 10^{14}\Msun$. At fixed mass, this reduces the number density of clusters by $\lesssim10\%$. Further following the redshift evolution, the effects of baryon depletion propagate up to more massive clusters. For low redshifts $z\lesssim0.3$, and for masses up to $\sim10^{14}\Msun$, our Hydro HMF is low by about $10-30\%$ compared to the DMonly case, and the difference increases when going from $\Delta_{200\textrm m}$ to $\Delta_{200\textrm c}$ to $\Delta_{500\textrm c}$. This dependence on the SO definition is consistent with the picture of baryons being ejected from the cluster central regions, as $r_{500\textrm c}<r_{200\textrm c}<r_{200\textrm m}$.

In the mass range between $10^{14}-10^{15}\Msun$, there is essentially no difference between our Hydro and DMonly simulations; as expected, baryonic effects are negligible on these scales. For even higher masses $>10^{15}\Msun$, and for $\Delta_{200\textrm m}$ and $\Delta_{200\textrm c}$, the Hydro simulations contain fewer clusters than DMonly. First, we check that our simulations have converged and correctly estimate the abundance of the most massive haloes. As discussed in Appendix~\ref{sec:initialization}, we run two additional DMonly realizations of Box1/mr, initialized at higher redshifts $z_\textrm{ini}=80$ and $z_\textrm{ini}=100$, and find that they agree very well with our default simulations initialized at redshift $z_\textrm{ini}=60$. Then, the high-mass end is much better constrained for Hydro where we also use data from Box0/mr which is not available for DMonly. We therefore expect that these differences would reduce when analyzing a larger simulation box for DMonly.

%%%%%

\subsection{Comparison with other dark matter only fits} \label{sec:MFcompare}

The comparison of our simulations and fits with Tinker08 and Watson13 can be summarized as follows:
\begin{enumerate}
\item At $\Delta_{200\textrm m}$, the fit by Tinker08 and our DMonly fit are in very good agreement. This is an important cross-check, since both simulations assume the same physical model and are similar in volume, and the assumed functional forms of the HMF are identical.
\item The fit by Watson13 tends to predict less low-mass haloes than the other HMFs considered here. At the high-mass end, for $\Delta_{200\textrm m}$, Watson13 predict less haloes than Tinker08 or our fits, but, for larger SO, Watson13 roughly agrees with Tinker08.
\item For the larger SO $\Delta_{200\textrm c}$ and $\Delta_{500\textrm c}$, Tinker08 predicts up to $\sim25\%$ more haloes for $M\lesssim10^{12}\Msun$; this difference steadily reduces to zero for masses $M\ge10^{14}\Msun$.
\end{enumerate}

Item (iii) above is due to a different approach in halo finding: We identify haloes and extract their masses using {\small SUBFIND}. The SO masses are constructed around the point corresponding to the minimum in the gravitational potential, neglecting substructure that eventually exceeds the overdensity $\Delta$ when going to high values like $\Delta_{500\textrm c}$. In contrast, Tinker08 use an SO finder that reveals these subhaloes. In Appendix~\ref{sec:substructure}, we confirm that the presence of substructure increases the abundance of objects by about $25\%$ on low mass scales $M\sim10^{12}\Msun$; the effect drops to zero for larger masses $M_{500\textrm c}\gtrsim10^{13}\Msun$ \citep[see also][]{kravtsov04}. Therefore, we expect this effect to be negligible for current cluster studies.

%%%%%

\subsection{Halo mass function fits} \label{sec:MFfits}

In Table~\ref{tab:MFparams}, we present the best fit parameters for the fits to the data from our DMonly and Hydro simulations. We also evaluate the $\chi^2$ per degree of freedom $\nu$ at the best-fitting location and the corresponding probability to exceed. The latter takes values that are very close to unity for all six HMF fits, indicating that the adopted form of the HMF is suitable for fitting the data.

The HMF parameters for $\Delta_{200\textrm m}$ can be directly compared with the literature because in this case the functional form of the HMF is identical. For DMonly, we find a value of the exponential cut-off scale $c=1.19\pm0.03$ that is fully consistent with Tinker08 ($c=1.19$), but significantly smaller than the $c=1.318$ in Watson13. Note that a large value of $c$ corresponds to a low cut-off scale in mass (see Equation \ref{eq:fsigma}), resulting in the underprediction of massive haloes by Watson13 as shown in Fig.~\ref{fig:dNdM_relative}. Finally, there is no evidence for redshift-evolution of the cut-off scale in the DMonly case with $c_z = -0.021\pm0.054$, but clear evidence in the case of the Hydro HMF, where $c_z = -0.045\pm0.012$.

The HMFs for $\Delta_{200\textrm c}$ and $\Delta_{500\textrm c}$ are often used in observational studies. For the latter SO, we also show the covariance matrix for the Hydro HMF parameters in Table~\ref{tab:covmatM500c}. This information should be used with the best-fitting parameters from Table~\ref{tab:MFparams} to capture the (statistical) uncertainties related to our HMF. For reference, the $1\sigma$ uncertainty on the HMF at $M_{500\textrm c}=10^{15}\Msun$ is $\sim1\%$.

%%%%%%%%%%%%%%%%%%%%%%%%%%%%%%%%%%%%%%%%%%%%%%%%

\begin{figure}
\includegraphics[width=\columnwidth]{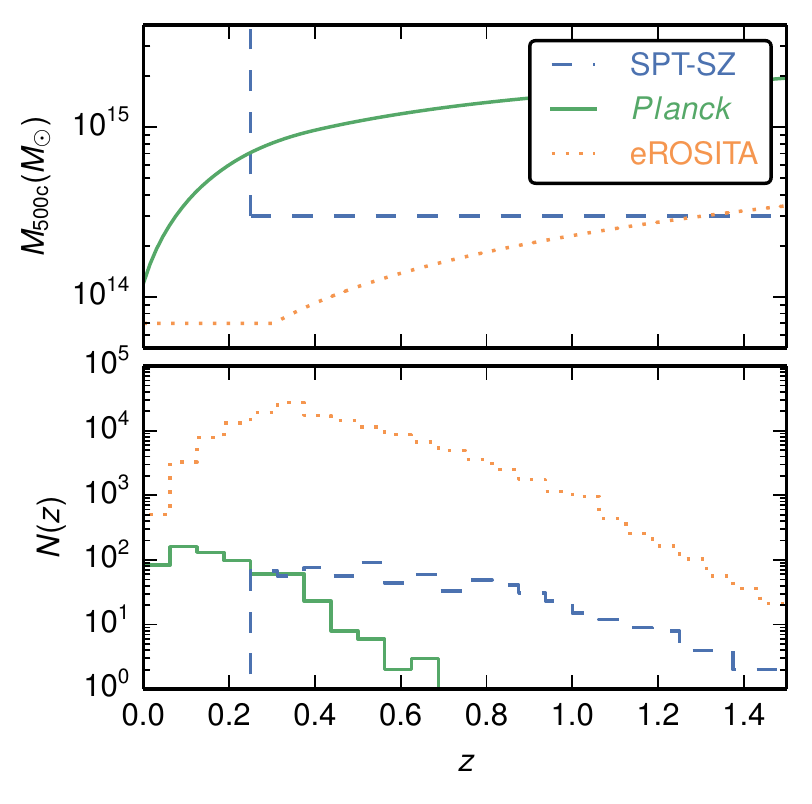}
\caption{Simulation of idealized representations of the SPT, \planck\, and eROSITA cluster catalogues. \textit{Top:} Selection functions used to create the catalogues. \textit{Bottom:} Distribution of simulated clusters as a function of redshift.}
\label{fig:surveylimits}
\end{figure}

\section{Cosmological impact} \label{sec:cosmoimpact}

There are differences between the HMFs extracted from our Hydro and DMonly simulations, and those from the literature. When used to interpret real cluster samples, the different HMFs will ultimately lead to different cosmological results. In the following, we quantify and discuss this effect. To this end, we create simulated cluster catalogues using our best-fitting Hydro HMF, and use either the Hydro, the DMonly, or literature fits to perform cosmological analyses. Because the baryonic impact on the HMF depends on mass and redshift, we expect qualitatively different shifts when using different HMFs, depending on the properties of a specific cluster survey.

We create and analyse three sets of simulated catalogues, whose properties approximately match real samples from the South Pole Telescope \citep[SPT;][]{carlstrom11}, the \planck\ satellite \citep{tauber10} and eROSITA \citep{predehl14}. The selection functions we assume are shown in Fig.~\ref{fig:surveylimits} and will be discussed in more detail. All samples are defined for SO $\Delta_{500\textrm c}$.

\begin{figure*}
\centering
\subfigure[Results for input $\Omega_\textrm m=0.272$.]{
	\includegraphics[width=\columnwidth]{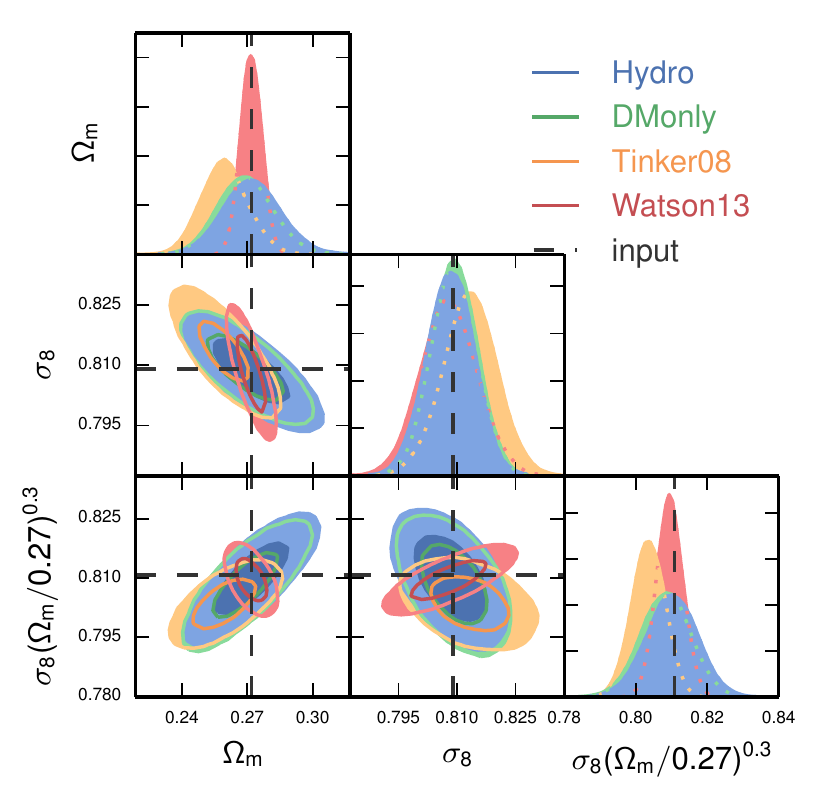}
	\label{fig:cosmo_SPT_027}}
\hspace{10pt}
\subfigure[Results for input $\Omega_\textrm m=0.22$.]{
	\includegraphics[width=\columnwidth]{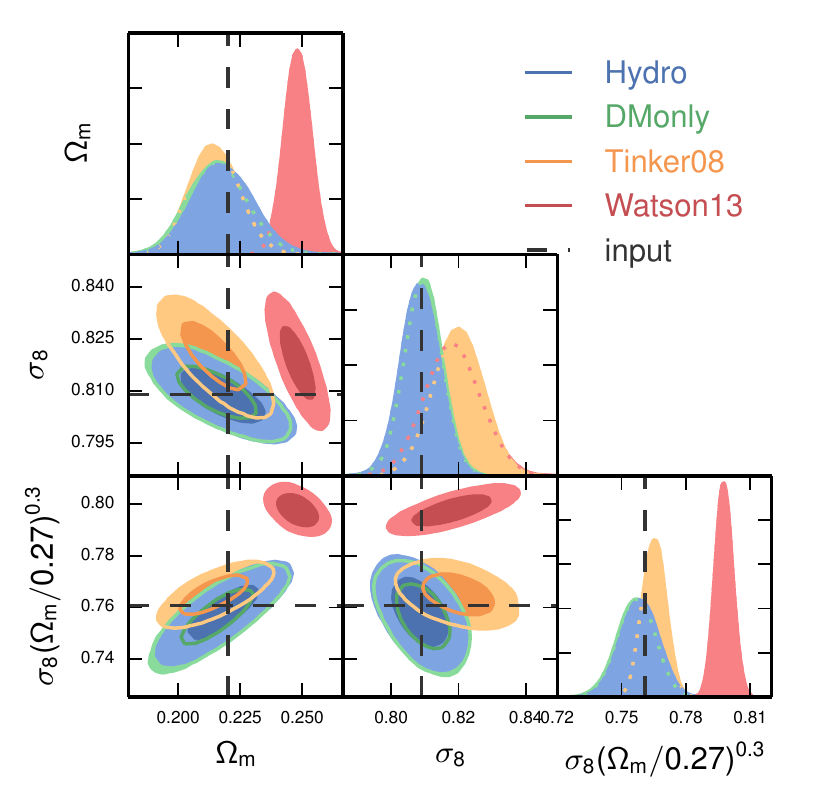}
	\label{fig:cosmo_SPT_022}}
\caption{Cosmological results from simulated SPT-like cluster samples, created using our Hydro halo mass function. The two different input cosmologies are marked by the dashed lines, in both cases $\sigma_8=0.809$. We show likelihood contours (68\% and 95\% confidence) in \Om-\sig-\sigOm\ space. Results obtained from different halo mass functions are colour-coded. The baryonic impact of the halo mass function is negligible for this high-mass and high-redshift sample. For the input value $\Omega_\textrm m = 0.272$, there is a small shift toward smaller values of \Om\ and \sigOm\ when using \citet{tinker08}. The constraints obtained from \citet{watson13} are tighter, but seem to be biased towards $\Omega_\textrm m\sim0.27$. See discussion in the text for more details.}
\label{fig:cosmo_SPT}
\end{figure*}

\begin{figure*}
\centering
\subfigure[Results for a simulated \planck-like cluster sample. The input cosmology is chosen to match the \planck\ CMB constraints \citep{planck15-13}. Given the large mass limit of the sample there is little difference between the results from our Hydro or DMonly halo mass functions. Using the \citet{tinker08} fit leads to shifts $\Delta(\sigma_8)\simeq-0.01$, and $\Delta($\sigOm$)\simeq-0.02$.]
	{\includegraphics[width=\columnwidth]{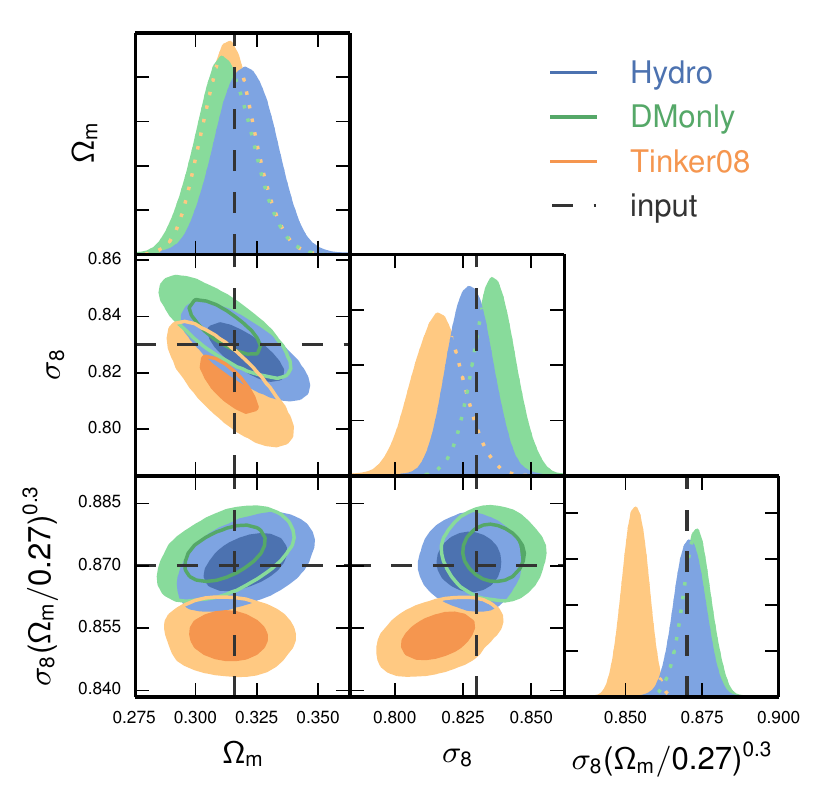}
	\label{fig:cosmo_Planck}}
\hspace{10pt}
\subfigure[Result for a simulated eROSITA-like cluster sample. For this sample, there is a difference between the results from the Hydro and the DMonly halo mass functions, which is due to the impact of baryons. Neglecting the baryonic impact on the halo mass function leads to an underestimate of $\Delta(\Omega_\textrm m)\simeq\Delta($\sigOm$)\simeq-0.01$. The DMonly and the \citet{tinker08} fits produce very similar constraints for this sample.]
	{\includegraphics[width=\columnwidth]{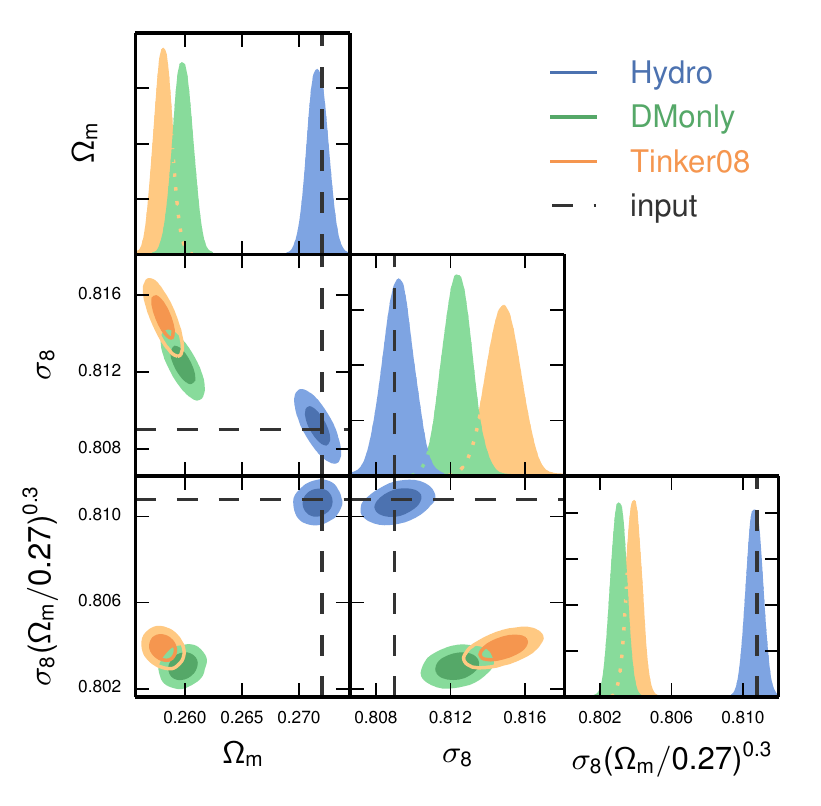}
	\label{fig:cosmo_eROSITA}}
\caption{Cosmological results from simulated realizations of approximations to the \planck\ and eROSITA cluster samples, created using our Hydro halo mass function. The input cosmologies are marked by the dashed lines. We show likelihood contours (68\% and 95\% confidence) in \Om-\sig-\sigOm\ space. Results obtained from different halo mass functions are colour-coded. Note the very different scales in both panels.}
\label{fig:cosmo_Planck_eROSITA}
\end{figure*}

We use the fit method described in Section~\ref{sec:analysis}, but we now fit for the cosmological parameters, and keep the HMF parameters fixed. Since no covariance matrix is available for the literature HMFs we compare to, we use our best-fit parameters without uncertainties, too, in order to make a comparison on equal footing. We further show that the (statistical) uncertainty on our Hydro HMF has negligible impact on the cosmological constraints. We restrict this analysis to the parameters \Om\ and \sig, which strongly affect the measured cluster abundance. Remember that these parameters enter the HMF calculation in Equation \ref{eq:MF} through their impact on the matter power spectrum $P(k,z)$ and the matter density $\bar \rho_\textrm m$. The \Om-\sig\ likelihood contours from the cluster number counts experiment exhibit a characteristic, elongated degeneracy in the \Om-\sig\ plane (see Figs~\ref{fig:cosmo_SPT} and \ref{fig:cosmo_Planck_eROSITA}). The parameter combination \sigOm\ is interesting because it reflects the width of this degeneracy, i.e. the direction in \Om-\sig\ space which is best constrained using clusters. We show the constraints we recover on this parameter combination, too.

In this test, we directly use the simulated cluster masses. That is, we do not include any systematic uncertainties and measurement errors related to mass estimation as one would have to do for a real cluster sample. This also means that the uncertainties we recover only represent the statistical uncertainties related to the sample size, and cannot be compared with results from observed clusters. The aim of this analysis is to investigate and quantify offsets related to the HMF, which justifies this simplified approach. For this same reason, we do not quote the errors on the recovered parameters.

The typical uncertainties on the cosmological parameters from current cluster samples are $\sigma (\Omega_\textrm m) \sim \sigma(\sigma_8) \sim \sigma($\sigOm$) \sim0.03$ \citep[e.g.][]{planck13-20,wtg14d}. We will refer to these characteristic numbers in the following.

%%%%%%%%%%

%%%%%%%%%%

\subsection{Cosmological analysis of an SPT-like cluster sample}

The SPT sample is selected through the cluster Sunyaev--Zel'dovich effect \citep[SZE;][]{sunyaev72} signature, and we approximate the catalogue as mass-selected with $M_{500\textrm c}>3\times10^{14}\Msun$, and restrict to redshifts $z>0.25$ \citep[see Fig.~\ref{fig:surveylimits} and cf.][]{bleem15}. For the SPT survey of size $2500$~deg$^2$, our simulated catalogue contains $687$ systems. We consider three different input cosmologies with different values of $\Omega_\textrm m = 0.22, 0.272, 0.4$ with the same $\sigma_8=0.809$ in each case. A subset of the results appears in Table~\ref{tab:cosmo} and Fig.~\ref{fig:cosmo_SPT}.

The results from both our Hydro and DMonly HMFs show nearly perfect agreement, indicating that the effect of baryons on the HMF is negligible in this case. This is expected, because, as previously noted, the impact of baryons is most important for low-mass clusters at low redshifts (see Section~\ref{sec:baryonimpact}). The SPT-like survey does not probe this mass and redshift regime.

The marginalized constraints on \sig\ obtained using Tinker08 are in very good agreement with the ones from our HMFs, but \Om\ slightly shifts by $-0.01$. Due to a mild tilt in degeneracy axes, \sigOm\ is slightly tighter when using the Tinker08 fit.

The constraints obtained on \Om\ using the Watson13 fit are tighter than the ones just discussed (see Fig.~\ref{fig:cosmo_SPT}). However, these results seem to be biased toward $\Omega_\textrm m\sim0.27$. For example, the preferred value recovered for the sample with input $\Omega_\textrm m=0.22$ is $\Omega_\textrm m (\textrm{Watson13})=0.25$, and we further obtain $\Omega_\textrm m (\textrm{Watson13})=0.31$ for an input value $\Omega_\textrm m=0.4$. Their assumed form of the redshift dependence of the fit parameters (Equation~\ref{eq:MF_z} in this work, Equations~13-15 in Watson13) involves $\Omega_\textrm m(z)$. We suspect that this parametrization introduces an implicit and spurious preference for $\Omega_\textrm m \sim0.27$, which is their simulation input value. We will not consider the Watson13 fit for the rest of this work.

The statistical uncertainty of our Hydro HMF is captured by the covariance matrix in Table~\ref{tab:covmatM500c}. We repeat the cosmological analysis using the Hydro HMF and its parameter covariances, and infer the additional uncertainties due to the uncertainty on the HMF $\sigma_\textrm{HMF}$ using quadrature addition. We find $\sigma_\textrm{HMF}(\Omega_\textrm m) \sim 0.004$ and $\sigma_\textrm{HMF}(\sigma_8) \sim 0.002$, and conclude that the statistical uncertainties on our HMF are completely negligible for current cluster samples.

%%%%%%%%%%

\begin{table}
\caption{Mean recovered cosmological parameter values from different halo mass functions for simulated cluster samples. We do not show errors because we are interested in absolute shifts only.}
\begin{tabular}{lccc}
Parameter & \Om\ & \sig\ & \sigOm \\\hline
Input & $0.272$ & $0.809$ & $(0.811)$\\\hline
\multicolumn{4}{l}{SPT-like sample}\\
Hydro & $0.272$ & $0.808$ & $0.810$ \\
DMonly & $0.269$ & $0.809$ & $0.808$ \\
\cite{tinker08} & $0.260$ & $0.813$ & $0.804$ \\\hline
\multicolumn{4}{l}{eROSITA-like sample}\\
Hydro & $0.272$ & $0.809$ & $0.811$\\
DMonly & $0.260$ & $0.812$ & $0.803$\\
\cite{tinker08} & $0.258$ & $0.815$ & $0.804$\\\hline
\multicolumn{4}{l}{\textit{Planck}-like sample}\\
Input & $0.316$ & $0.830$ & ($0.870$)\\[3pt]
Hydro & $0.320$ & $0.828$ & $0.871$\\
DMonly & $0.312$ & $0.836$ & $0.873$\\
\cite{tinker08} & $0.314$ & $0.816$ & $0.853$\\\hline
\end{tabular}
\label{tab:cosmo}
\end{table}

%%%%%%%%%%

%%%%%%%%%%

\subsection{Cosmological analysis of a \planck-like cluster sample}

The \planck\ cluster sample is selected using the SZE, too, and extends down to redshift $z=0$. However, the satellite's beam is larger than the SPT beam, and the survey mass limit varies significantly with redshift. We mimic the \planck\ selection function following the sample mass-redshift distribution shown in Fig.~1 in \cite{planck15-24}. We further assume a hydrostatic bias of $1-b\simeq0.8$, which then leads to the selection function we show in Fig.~\ref{fig:surveylimits}. For this exercise, we choose our input cosmology to match the values preferred by the \planck\ CMB anisotropy measurement \citep[$\Omega_\textrm m=0.316$, $\sigma_8=0.83$, $H_0 = 67.3$~km~s$^{-1}$~Mpc$^{-1}$;][]{planck15-13}. Assuming a sky coverage of 65\%, the simulated catalogue contains $639$ clusters.

The results are shown in Table~\ref{tab:cosmo} and Fig.~\ref{fig:cosmo_Planck}. We recover very similar constraints on \Om\ for our Hydro and DMonly HMFs and Tinker08. There is some spread in the constraints on \sig, where the difference between our DMonly and Tinker08 is $\Delta(\sigma_8)\simeq0.02$. We recover identical constraints on \sigOm\ from our Hydro and DMonly fits, but using Tinker08 leads to an shift $\Delta($\sigOm$)\simeq-0.02$. The shift in \sig\ and the induced offset in \sigOm\ are larger than the differences in our analysis of the SPT-like sample. As discussed in Section~\ref{sec:MFcompare}, our HMFs and Tinker08 differ somewhat at the highest clusters masses for redshifts $z\geq0.3$. Then, as shown by the selection function in Fig.~\ref{fig:surveylimits}, the \planck-sample contains more massive clusters than the SPT-sample which makes the \planck-sample more sensitive to differences in the high-mass end of the HMF.

The analysis of the real \planck\ cluster sample is limited by the systematic uncertainty on the overall mass scale, often parametrized by the mass bias factor. In their latest cluster cosmology analysis, the \planck\ collaboration adopts measurements of the mass bias from different authors \citep{planck15-24}. Assuming a measurement of the mass bias from ground-based weak lensing, the recovered values of \sig\ are low by roughly $0.05-0.08$ compared to the value recovered from the \planck\ CMB anisotropy measurements. However, due to the combination of systematic and statistical uncertainties, the \planck\ CMB preferred value in \Om-\sig\ space is still compatible with the cluster measurement at the $\sim1-2\sigma$ level. Our simplified analysis suggests that using our HMF instead of Tinker08 could lead to a shift in \sigOm\ by about $+0.02$, which would reduce the differences between the values preferred by \planck\ clusters and the CMB anisotropy.

%%%%%%%%%%

\subsection{Cosmological analysis of an eROSITA-like cluster sample}

The eROSITA cluster sample will be X-ray flux selected and extend from redshift $z=0$. For the present test, we assume a detection limit of 50 photons in the $0.5-2.0$ keV band with a typical exposure time of 1.6 ks. We model this selection as a combination of a redshift-dependent mass threshold $M_{500\textrm c}>2.3\,z\times10^{14}\Msun$, with an additional mass cut at $7\times10^{13}\Msun$ (see Fig.~\ref{fig:surveylimits} and compare with fig.~2 in \cite{pillepich12}, and also \cite{merloni12,borm14}). The eROSITA full-sky catalogue simulated in this way contains $\sim1.45\times10^5$ clusters.

The results of the analysis of this sample appear in Table~\ref{tab:cosmo} and Fig.~\ref{fig:cosmo_eROSITA}. The recovered constraints are very tight due to the large cluster sample and the fact that we do not include mass measurement uncertainties. For this sample, the constraints on \sig\ agree to within $\Delta(\sigma_8)\simeq0.006$. More importantly, the recovered constraints on \Om\ are in good agreement between our DMonly and Tinker08, but both are low compared to the results obtained using the Hydro HMF. This is an indication that baryonic effects are indeed important for this sample. As previously discussed, baryons have their strongest impact on the HMF at low redshifts and for low masses, which is a regime that is well probed by eROSITA. Therefore, neglecting the baryonic impact for this sample and using a dark matter only HMF would lead to a bias of $\Delta\Omega_\textrm m \simeq-0.01$. This bias is of the same order as the expected uncertainty from eROSITA $\sigma(\Omega_\textrm m)\simeq0.012$ \citep{pillepich12}, meaning that the impact of baryons on the HMF will have to be accounted for in the cosmology analysis. For the same reasons just discussed, \sigOm\ is slightly underestimated when using either our DMonly or the Tinker08 fit.

%%%%%%%%%%%%%%%%%%%%%%%%%%%%%%%%%%%%%%%%%%%%%%%%

\section{Summary} \label{sec:summary}

We calibrate the HMF and investigate the impact of baryons using the hydrodynamic \magneticum\ simulations together with dark matter only counterparts. Our simulations and the halo selection are characterized by (1) a treatment of the baryonic component and of AGN feedback that is in good agreement with several observations such as baryon fractions, AGN luminosity functions \citep{hirschmann14} and cluster pressure profiles \citep{planck13-5,mcdonald14}, (2) large cosmological volumes probed by boxes of up to $(3818~\textrm{Mpc})^3$, which allow us to track cluster masses up to a $M\gtrsim10^{15}\Msun$, and (3) a conservative halo selection with $>10^4$ dark matter particles within $r_\Delta$, minimizing potential biases related to numerical resolution. To avoid a different sampling of the initial density fluctuations, the DMonly simulations were run using two species of dark matter with masses corresponding to those of the dark matter and baryonic particles in the Hydro simulations. We extract SO masses $\Delta_{200\textrm m}$, $\Delta_{200\textrm c}$, and $\Delta_{500\textrm c}$.

The presence of baryons tends to decrease the cluster masses, which -- given the shape of the HMF -- leads to a decrease of the expected number of objects for a given mass (see Figs~\ref{fig:dNdM} and \ref{fig:dNdM_relative}). The number density of haloes decreases by up to $\sim15\%$ for low masses $M\lesssim10^{14}\Msun$ and at low redshifts $z\lesssim0.5$. At higher masses and redshifts, our Hydro and DMonly simulations agree very well. Qualitatively similar results have been recently presented in other publications \citep{cui14,cusworth14,velliscig14,vogelsberger14,schaller15}. \citet{martizzi14} find a mild increase of the HMF due to baryons. 

The HMF shape varies only weakly with redshift \textit{and} cosmology when masses are defined either by FoF with $b\simeq0.2$, or for SO $\Delta_\textrm{mean}\sim200$. Therefore, for the HMF for $\Delta_{500\textrm c}$ ($\Delta_{200\textrm c}$), we introduce a mapping between $M_{500\textrm c}$ ($M_{200\textrm c}$) and $M_{200\textrm m}$ as a function of mass, redshift, and \Om, and argue that this allows us to use the universal properties of $\Delta_{200\textrm m}$ also for masses defined by $\Delta_{500\textrm c}$ ($\Delta_{200\textrm c}$). In practice, our HMF fits are used as follows:
\begin{itemize}
\item Calculate $\sigma(M,z)$ using Equation~\ref{eq:sigma}.
\item Calculate $f(\sigma,z)$ from Equations \ref{eq:fsigma} and \ref{eq:MF_z}, using the appropriate parameters from Table~\ref{tab:MFparams}.
\item For $\Delta_{200\textrm m}$: The HMF is given by Equation \ref{eq:MF}.
\item For $\Delta_{500\textrm c}$: Calculate $M_{500\textrm c} / M_{200\textrm m}$ from Equations \ref{eq:dMvirdM} and \ref{eq:lastM500M200}, and obtain the HMF from Equation~\ref{eq:MF500c}.
\item For $\Delta_{200\textrm c}$: Calculate $M_{200\textrm c} / M_{200\textrm m}$ from Equations~\ref{eq:M200cM200m} -- \ref{eq:M200c_Om}, and obtain the HMF using Equation~\ref{eq:MF200c}.
\end{itemize}
Note that the same approach could be used to propagate the universal behaviour of the $\Delta_{200\textrm m}$ HMF to any overdensity $\Delta$.

We investigate how the differences among our Hydro, DMonly and previously published dark matter only HMFs affect cosmological results from cluster abundance measurements. To this end, we simulate idealized representations of the SPT, \planck, and eROSITA surveys, assuming simplified selection schemes as shown in Fig.~\ref{fig:surveylimits}. We assume perfect knowledge of cluster masses $M_{500\textrm c}$, and do not account for any uncertainties or systematics related to mass-observable relations. Therefore, the cosmological parameter uncertainties we recover here are tighter than the actual constraints that would be obtained in a comprehensive analysis of real data. Moreover, neglecting the conversion from observable to mass likely removes some cosmological dependencies. However, this test can be used as guidance in understanding the impact of differences in the HMF.

The results of these analyses can be summarized as follows (see also Figs~\ref{fig:cosmo_SPT} and \ref{fig:cosmo_Planck_eROSITA} and Table \ref{tab:cosmo}):
\begin{itemize}
\item For the SPT-like and \planck-like samples, the impact of baryons is negligible, and we obtain identical cosmological results using either our Hydro or DMonly HMFs.
\item For the SPT-like sample, results obtained using the Tinker08 fit essentially agree with results obtained using our HMFs.
\item The HMF by Watson13 seems to bias results toward $\Omega_\textrm m\sim0.27$. This may be due to their parametrization of the redshift evolution of the HMF shape parameters using $\Omega_\textrm m(z)$, which results in a heightened and likely artificial cosmological sensitivity.
\item For the \planck-like sample, using our Hydro HMF instead of Tinker08 shifts the results by $\Delta($\sigOm$)\simeq0.02$. This shift corresponds to about half the observed difference between the latest \planck\ clusters and CMB constraints; using our HMF should therefore lead to better agreement between the two probes.
\item The eROSITA sample extends to lower masses than the SPT and \planck\ catalogues.  We observe an offset in the results from Hydro and DMonly, which we identify as the impact of baryons. Neglecting this effect leads to an underestimate $\Delta\Omega_\textrm m\simeq\Delta($\sigOm$)\simeq -0.01$, which is comparable to the expected overall uncertainty.
\end{itemize}

Part of the differences between the cosmological results recovered using our HMFs and using the Tinker08 could be due to different interpretations of the connection between the observed galaxy \textit{clusters} and their representation as \textit{haloes} in the simulations. Different assumptions lead to different halo-finding approaches: We extract SO masses around the minimum in the gravitational potential in each halo using {\small SUBFIND}. In contrast, Tinker08 employ an SO finder, and also include subhaloes in their fit. These different approaches only affect the HMF for low masses $M_{500\textrm c}\lesssim10^{13}\Msun$ and are therefore unimportant for the cosmological study of current cluster samples.

More work, both on the theoretical and on the numerical aspects of calibrating the HMF is needed to be able to fully extract the cosmological information from near-future cluster samples. It is important to better understand the cosmological dependencies of the fitting functions, and to construct an analytic formula whose universality -- or indeed departure from universality -- is well understood. Finally, a careful comparison of cluster catalogues generated from different sets of numerical simulations would be helpful to better understand the systematic uncertainties on the HMF.

%%%%%%%%%%%%%%%%%%%%%%%%%%%%%%%%%%%%%%%%%%%%%%%%

\section*{Acknowledgements}
We acknowledge the support of the DFG Cluster of Excellence ``Origin and Structure of the Universe'' and the Transregio programme TR33 ``The Dark Universe''. The calculations have partially been carried out on the computing facilities of the Computational Center for Particle and Astrophysics (C2PAP) and of the Leibniz Supercomputer Center (LRZ) under the project IDs pr58we, pr83li, and pr86re. Special thanks go to LRZ for the opportunity to run the Box0 simulation within the Extreme Scale-Out Phase on the new SuperMUC Haswell extension system. We appreciate the support from the LRZ team, especially N. Hammer, when carrying out the Box0 simulation.

\bibliography{magneticum}

%%%%%%%%%%%%%%%%%%%%%%%%%%%%%%%%%%%%%%%%%%%%%%%%

\appendix
\section{Spherical overdensity $\Delta_{200\textrm c}$}
\label{sec:SO200c}

Applying the method described in Section~\ref{sec:SO}, the HMF for $M_{200\textrm c}$ is
\begin{equation} \label{eq:MF200c}
\frac{dn}{dM_{200\textrm c}} = f(\sigma)\frac{\bar\rho_\textrm m}{M_{200\textrm c}}\frac{d\ln\sigma^{-1}}{dM_{200\textrm c}}\times\frac{M_{200\textrm c}}{M_{200\textrm m}}.
\end{equation}
Assuming a cluster density profile \citep{navarro97} and a mass-concentration relation \citep{duffy08}, we establish a mass-dependent fit for $M_{200\textrm c}/M_{200\textrm m}$
\begin{equation} \label{eq:M200cM200m}
\frac{M_{200\textrm c}}{M_{200\textrm m}} \equiv \gamma + \delta \ln \frac{M_{200\textrm c}}{\Msun},
\end{equation}
where $\gamma$ and $\delta$ depend on \Om\ and redshift as
\begin{align} \label{eq:M200c_z}
\gamma(\Omega_\textrm m,z) &= \gamma_0+\gamma_1\exp\left(-\left(\frac{\gamma_2-z}{\gamma_3}\right)^2\right) \nonumber\\
\delta (\Omega_\textrm m,z) &= \delta_0+\delta_1 z
\end{align}
and
\begin{align} \label{eq:M200c_Om}
\gamma_0 (\Omega_\textrm m) &= 3.54\times10^{-2} + \Omega_\textrm m ^{0.09} \nonumber\\
\gamma_1 (\Omega_\textrm m) &= 4.56\times10^{-2} + 2.68\times10^{-2}/\Omega_\textrm m \nonumber\\
\gamma_2 (\Omega_\textrm m) &= 0.721 + 3.50\times10^{-2}/\Omega_\textrm m \nonumber\\
\gamma_3 (\Omega_\textrm m) &= 0.628 + 0.164/\Omega_\textrm m \nonumber\\
\delta_0 (\Omega_\textrm m) &= -1.67\times10^{-2} + 2.18\times10^{-2}\,\Omega_\textrm m \nonumber\\
\delta_1 (\Omega_\textrm m) &= 6.52\times10^{-3} - 6.86\times10^{-3}\,\Omega_\textrm m.
\end{align}
This fit is accurate at the few percent level in the range $0<z<2$, $10^{13}<M_{200\textrm c}/\Msun<2\times10^{16}$, and $0.15<\Omega_\textrm m<0.5$.

%%%%%%%%%%

\begin{figure}
\includegraphics[width=\columnwidth]{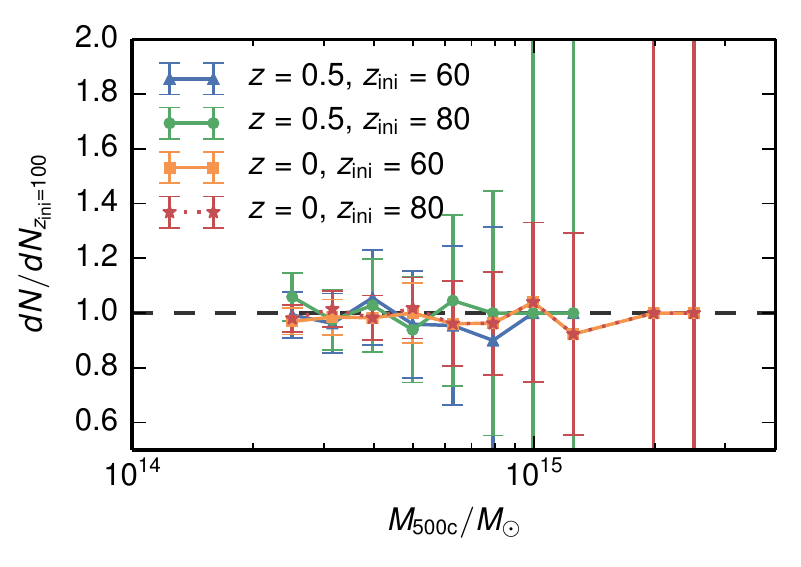}
\caption{Number of haloes extracted at redshift $z=0.5$ and $z=0$ from simulations started at initial redshift $z_\textrm{ini}=60$ (used in this work) and two control simulations started at higher initial redshifts $z_\textrm{ini}=80$ and $z_\textrm{ini}=100$. The error bars show Poisson uncertainties. All three simulations agree within the statistical uncertainties, indicating good convergence of the simulations used in this work.}
\label{fig:z_ini}
\end{figure}

\section{Impact of initial redshift} \label{sec:initialization}

The simulations used in this work are initialized at redshift $z_\textrm{ini}=60$ using the Zel'dovich approximation. Under this assumption, running a simulation from too low an initial redshift can lead to the suppression of the formation of high-mass haloes \citep{reed13}.

We confirm that our simulations are converged by running two additional control simulations. These correspond to Box1/mr DMonly as used in the main body of this work, except that we choose higher initial redshifts, $z_\textrm{ini}=80$ and $z_\textrm{ini}=100$. In Fig.~\ref{fig:z_ini}, we compare the abundance of haloes in the three simulations at redshifts $z=0.5$ and $z=0$. All three simulations agree very well within the error bars. We conclude that our simulations initialized at redshift $z_\textrm{ini}=60$ are converged and suitable for calibrating the HMF.

%%%%%%%%%%

\begin{figure}
\includegraphics[width=\columnwidth]{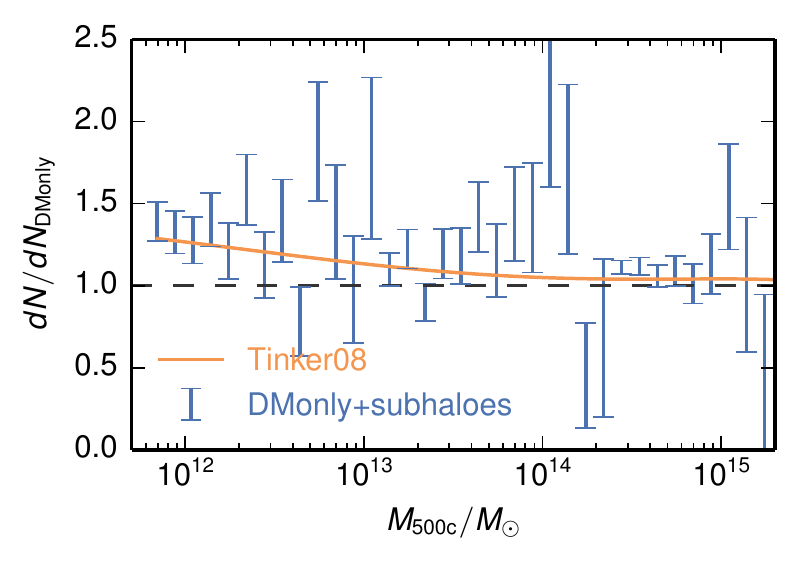}
\caption{Impact of substructure on the DMonly HMF for $\Delta_{500\textrm c}$. We show the abundance of haloes, including subhaloes, relative to the number of main haloes as used throughout this work. For small halo masses, substructure has a significant contribution to the abundance, and the abundance of haloes including substructure seems to be in good agreement with \citet{tinker08}. The impact of subhaloes vanishes for haloes with $M_{500\textrm c}\gtrsim10^{13}\Msun$.}
\label{fig:exposed}
\end{figure}

\section{Substructure at $\Delta_{500\textrm c}$} \label{sec:substructure}

As discussed in Section~\ref{sec:MFcompare}, the Tinker08 HMF predicts more haloes for SO $\Delta_{200\textrm c}$ and $\Delta_{500\textrm c}$ at low masses $M\lesssim10^{14}\Msun$ than our DMonly fit, and both HMF agree much better for $\Delta_{200\textrm m}$. This is mainly due to different assumptions about halo definition and identification. In this section, we show that our simulations are in good agreement with the low-mass end of the Tinker08 HMF at $\Delta_{500\textrm c}$ when haloes are extracted assuming the same halo definition as applied in Tinker08. The different definitions are:
\begin{itemize}
\item In this work, we extract spherical halo masses using the minimum in the gravitational potential within each halo as the center. In this approach, a halo with a given mass for one SO definition has exactly one counterpart in any other SO definition. 
Another consequence is that no subhalo that would actually exceed large overdensity thresholds like $\Delta_{500\textrm c}$ is extracted within a halo.
\item In Tinker08 SO masses are computed for all (potentially overlapping) haloes whose centers are separated by more than $r_\Delta$. As a consequence, for high overdensities, and within massive haloes, particles in the main halo and in its subhaloes may be counted multiple times. Furthermore, a halo identified at some SO can correspond to multiple exposed haloes at higher overdensity.
\end{itemize}

We investigate the impact of substructure at $\Delta_{500\textrm c}$ in our simulations by extracting another sample of haloes adopting the Tinker08 approach. We apply a SO finder to our DMonly simulations at redshift $z=0$ and extract all haloes that exceed the SO $\Delta_{500\textrm c}$; this sample also includes subhaloes. In Fig.~\ref{fig:exposed}, we compare the abundance of objects extracted in this way to the abundance of main haloes as extracted by {\small SUBFIND} and used throughout this work. Both agree for masses above $\sim10^{13}\Msun$, but at lower masses, subhaloes contribute to the abundance by up to $\sim25\%$ for $M_{500\textrm c}\simeq10^{12}\Msun$. This observation is in agreement with the discussion in Tinker08 and \cite{kravtsov04}. Indeed, Fig.~\ref{fig:exposed} suggests that, for masses below $\sim10^{13}\Msun$, the total abundance of haloes including subhaloes extracted from our simulations agrees with the Tinker08 HMF. 

We conclude that, in the low-mass regime, the differences between the HMFs presented in this work and the HMF by Tinker08 are due to different assumptions on the halo definition. We note that these differences do not affect the shape of the HMF at higher masses $M\gtrsim10^{14}\Msun$ which are important for the cosmological implications discussed in this work. However, this also makes it clear that the choice of (sub)halo identification applied to any observed cluster sample must be consistent with the identification method applied to the simulation data. For the cluster samples observed with the SPT and \planck, these differences are negligible, but they may indeed be important when using groups selected by, for example, eROSITA.

\end{document}